\documentstyle[12pt]{article}

\begin{document}
\title{Radar equations in the problem of radio wave backscattering during bistatic
soundings}
\author{O.I.Berngardt, A.P.Potekhin}
\maketitle
\begin{abstract}
This paper outlines a method of obtaining the relation between the singly scattered
signal and the Fourier-spectrum of medium dielectric permititvity fluctuations, 
with due regard for the fact that the scattering volume is determined
by antenna patterns and is not small. On the basis of this equation we obtained
the radar equation relating the scattered signal spectrum to the spatial spectrum
of fluctuations. Also, a statistical radar equation is obtained, which relates
the mean statistical power of the scattered signal to the spectral density of
the dielectric permitivity fluctuations without a classical approximation of
the smallness of the irregularities spatial correlation radius. 
The work deals with the bistatic sounding case, when
the exact forward and exact backward scattering are absent, and sounding signal
have sufficiently norrow spectral band for scattered volume to change slowly on
ranges of Fresnel radius order. 
The statistical radar equations obtained differs from
the classical ones in the presence the coherent structures with big correlation
radii, and so the received signal spectrum can differ from intrinsic spectrum
of irregularities.
\end{abstract}

\section{Introduction}

The method of radio wave backscattering due to dielectric permitivity fluctuations
of the medium provides the basis for a wide variety of techniques for probing
the ionosphere (the radio waves incoherent scattering method \cite{Suni},
scattering from artificial irregularities\cite{Vilensky}), the atmosphere (mesospheric-
stratospheric-tropospheric sounding \cite{Woodman}), and other media. 
Central to these techniques is the radar equation that relates the mean spectral power 
of the scattered signal to the statistical characteristic of the meduium dielectric 
permitivity  fluctuations, their spectral density \cite{Tatarsky}, \cite{Woodman},
\cite{Isimaru}. A standard method for constructing the statistical radar equation
involves constructing the spectral power (or an received signal autocorrelation
 function) using two approximations. One of them, viz. the single-scattering
approximation, is applicable when the dielectric permitivity fluctuations are
weak and the scattered field is significantly weaker than the incident field.
The other approximation is the approximation of the irregularities spatial 
correlation radius smallness in comparison with the Fresnel radius. 

The single scattering problem of the electromagnetic wave has been reasonably
well-studied in situations were the receiver and the transmitter are in the
far-field region of the scatterer. In this case it is possible to obtain a simple
linear relation between the scattered signal and the spatial Fourier-spectrum
of irregularities without recourse to a statistical averaging \cite{Newton},
\cite{Isimaru}. However, in remote diagnostics of media, the scattering
volume size is determined by the antenna patterns crossing. Hence it is impossible
to use a classical approximation of the sounding volume smallness\cite{Isimaru}
to obtain the radar equation relating the scattered signal to the Fourier-spectrum
of dielectric permitivity fluctuations. Therefore, when obtaining the statistical
radar equation, one has to average the received signal power and to use
the approximation of the spatial correlation radius smallness, which
makes it possible to generalize results derived from solving a classical problem
of wave scattering from a single small irregularity to the problem of scattering
from a set of uncorrelated small irregularities \cite{Tatarsky},\cite{Isimaru}. 

Here we have obtained the radar equation relating the scattered signal
spectrum to the spatial spectrum of dielectric permitivity fluctuations for
bistatic sounding. This expression was obtained without a traditional
limitation on the smallness of the scattering volume size and describes
essentially the scattering from an extended scatterer. An analysis of the 
expression revealed a selectivity of the scattering similar to the 
widely-known Wolf-Bragg condition. 

We obtained the statistical radar equation for bistatic sounding
which relates the scattered signal mean power to the dielectric permitivity 
fluctuations spectral density without a classical approximation of
the spatial correlation radius smallness. This equation has a more extensive
validity range when compared to the well-known equation \cite{Tatarsky},\cite{Isimaru},
obtained by assuming the smallness of the irregularities spatial correlation radii. 

There were pointed out earlier \cite{DoviakZrnic} there is difficulties in 
obtaining the equation for correlation radii or scattering volume
bigger than Fresnel radius without exact antenna patterns, irregularities and
sounder signals. The suggested method do not contain those limitations.

\section{Starting equation }

Consider a bistatic sounding within the single-scattering approximation. We
use, as the starting equation, the well-known expression for a complex envelope
of the received signal \cite{Tatarsky}\cite{Isimaru} (accurate
to terms unimportant for the subsequent discussion): 

\begin{equation}
\label{eq:1}
u(t)=\int H(t,\overrightarrow{r})g(\overrightarrow{r})\frac{\epsilon (t-R_{s}/c,\overrightarrow{r})}{R_{s}R_{i}}exp(ik_{0}(R_{s}+R_{i}))d\overrightarrow{r}
\end{equation}

where 

\begin{equation}
\label{eq:2}
H(t,\overrightarrow{r})=o(t)a(t-(R_{i}+R_{s})/c)
\end{equation}

\( \overrightarrow{R_{i,s}}=\overrightarrow{r}-\overrightarrow{r_{i,s}} \)
- is the position of the point under investigation relative to \( \overrightarrow{r_{i,s}} \)
(the locations of the transmitter and the receiver); \( o(t) \) and \( a(t) \)
are, respectively, the time window of reception, and the emitted signal complex envelope, 
with the window and the signal being narrow-band and have the
bands \( \Delta \Omega _{o},\Delta \Omega _{a}<<\omega _{0} \) , \( \omega _{0} \)
is the carrier frequency,\( k_{0}=\omega _{0}/c \); \( \widehat{R}=\overrightarrow{R}/R \)
is a unit vector in a given direction; \( g(\overrightarrow{r})=f_{i}(\widehat{R_{i}})f_{s}(\widehat{R_{s}})\overrightarrow{l_{s}}\cdot [\widehat{R_{s}}\times [\widehat{R_{s}}\times \overrightarrow{l_{i}}]] \)
is the beam factor, where \( f_{i}(\widehat{r}) \) and \( f_{s}(\widehat{r}) \)
are the patterns of the transmit and receive antennas, \( \overrightarrow{l_{i}} \)
and \( \overrightarrow{l_{s}} \) are their polarization factors; and \( \epsilon (t,\overrightarrow{r}) \)
represents dielectric permitivity fluctuations. The expression (\ref{eq:1})
is obtained on the assumption that the scattering volume is in the far-field
range of the receive and transmit antennas \( R>>D^{2}/\lambda _{0} \), where
\( D \) is a typical antenna size, and \( \lambda _{0} \) is the wavelength
of the emitted signal. 

Equation (\ref{eq:1}) defines the relation between the received signal and
medium fluctuations and is essentially (in the sounding problem) the
radar equation for signals unlike the classical one for power characteristics.
The kernel \( H \) and the beam factor \( g \) are determined by transmitting 
and receiving system parameters. Specifically, the kernel \( H \)
determines the region of fluctuations \( \epsilon  \) over time and space which
contributes to the scattered signal. The beam factor \( g \) determines the
region of space which contributes to the scattered signal.

\section{Relation of the scattered signal to the spatial Fourier-spectrum of irregularities.}

\subsection{Derivation of radar equations.}

This section discusses the method of obtaining the relation between the 
scattere signal spectrum and the spatial spectrum of the medium dielectric 
permitivity fluctuations. The problem of obtaining a relation for the 
case of small scatterers was considered in \cite{Newton}; in this 
paper we have obtained a relation without a limitation on the size of the
object being probed. 
The main idea of this method is the transition from the problem of
scattering on spatial irregularities to the scattering on separate spatial
Fourier-harmonics of these irregularities. Their respective contributions are
calculated by the stationary-phase method and are summarized. 

Let us analyze the expression (\ref{eq:1}), without performing a standard \cite{Tatarsky} \cite{Isimaru} 
transition to quadratic (in field) characteristics. Within Born's approximation,
the main physical mechanism for signal shaping is the scattering on certain
Fourier-harmonics \( \epsilon  \). It is therefore convenient first to highlight
the relation between the scattered signal spectrum and spectral characteristics of
the medium. 

By going to the spectral representations for \( u \) and \( \epsilon  \)
in (\ref{eq:1}) we obtain following expression for received signal spectrum,
fully equaivalent to the initial (\ref{eq:1}):

\begin{equation}
\label{eq:3}
u(\omega )=\int I(\omega ,\nu ,k_{0},\overrightarrow{k})\epsilon (\nu ,\overrightarrow{k})d\nu d\overrightarrow{k}
\end{equation}

where the integral \( I \) is determined by
\begin{equation}
\label{eq:4}
I(\omega ,\nu ,k_{0},\overrightarrow{k})=\int \frac{H(\omega -\nu ,\overrightarrow{r})g(\overrightarrow{r})exp(i\varphi (\overrightarrow{r},k_{0},\overrightarrow{k}))}{R_{s}R_{i}} d\overrightarrow{r}
\end{equation}

\begin{equation}
\label{eq:5}
\varphi (\overrightarrow{r},k_{0},\overrightarrow{k})=\overrightarrow{k}\overrightarrow{r}+(k_{0}+\nu /c)R_{s}+k_{0}R_{i}
\end{equation}

The integral in (\ref{eq:4}) is proportional to the amplitude of the signal
scattered on a separate spatial harmonic and contains a rapidly oscillating
function. 

The virtue of (\ref{eq:3}) is that one can interpret received signal
without averaging, when we have the model of \( \epsilon (\nu ,\overrightarrow{k}) \),
because usually we have only spectral form. The drawbacks of such
representation (neglecting rapidly oscillating function under integral
(\ref{eq:4})) are that it is selective Bragg scattering character, 
which is well known for the small scatterers(\ref{eq:4}).
That is why we will transform this representation (\ref{eq:3}) to the form
with this drawback removed.

According to \cite{Newton} \cite{Tatarsky}, let us assume that
the main mechanism is the Bragg scattering and that the largest contribution
to the scattered signal is made by medium spatial harmonics, the wave
number of which has the order of twice the incident wave number and so will 
be large. The distances range from which the signal arrives
(determined by the crossing of the beam factor \( g \) and the weighting function
\( H \)), is in the antennas far-field range and also will be large. 
Therefore,the phase in (\ref{eq:5}), the product of the wave number by the 
distance, is a large parameter \( kr>>(D/\lambda_{0})^{2}>>1 \) wich makes 
it possible to evaluate this integral by the stationary-phase method(SPM) 
\cite{Fedoruk}, whose applicability conditions will be discussed below. 

The expression under the integral sign in (\ref{eq:4}) has a stationary point
\( \overrightarrow{r_{0}}(\overrightarrow{k}) \) which makes the main contribution
to the integral. Its location is defined by the equation 

\begin{equation}
\label{eq:6}
\overrightarrow{k}(\overrightarrow{r_{0}})=-((k_{0}+\nu /c)\widehat{R_{s}}(\overrightarrow{r_{0}})-k_{0}\widehat{R_{i}}(\overrightarrow{r_{0}})),
\end{equation}

which is a modified Wolf-Bragg condition for the scattering from nonstationary
spatial arrays. 

As a first approximation, the integral in (\ref{eq:4}) is therefore equal to
the contribution from the stationary point: 

\begin{equation}
\label{eq:7}
I(\omega ,\nu ,k_{0},\overrightarrow{k})=\frac{H(\omega -\nu ,\overrightarrow{r_{0}})g(\overrightarrow{r_{0}})}{R_{s}(\overrightarrow{r_{0}})R_{i}(\overrightarrow{r_{0}})}V(\overrightarrow{r_{0}})(2\pi )^{3/2}exp(i\varphi (\overrightarrow{r_{0}},k_{0},\overrightarrow{k})+i\pi /4)
\end{equation}

where 

\begin{equation}
\label{eq:8}
V(\overrightarrow{r})=\left[ det\left( \frac{d^{2}\varphi (\overrightarrow{r})}{dr_{i}dr_{j}}\right) \right] ^{-1/2}=\frac{R^{2}_{s}R^{2}_{i}}{k^{3/2}_{0}(1+\xi )|\overrightarrow{R_{s}}\times \overrightarrow{R_{i}}|(R_{s}(1+\xi )+R_{i})^{1/2}}
\end{equation}

where \( \xi =\nu /(ck_{0}) \) is usually small because of the narrow-bandedness
of the received signal. 

The quantity \( V \) has the meaning of a 'cophasal' region spatial 
volume that makes the main contribution to the amplitude of scattering on the
spatial harmonic in (\ref{eq:4}). Thus the main contribution to the integral
in (\ref{eq:4}) comes from 'cophasal' regions of spatial harmonic arrays having
a nearly ellipsoidal shape. The direction toward these regions is determined
by a modified Wolf-Bragg condition (\ref{eq:6}), and their volume is defined
by (\ref{eq:8}). It is evident that the linear dimension of the 'cophasal'
region has the Fresnel radius order \( R_{F}=(\lambda r)^{1/2} \). 

The scattered signal is the superposition of contributions from separate spatial
harmonics. Therefore, by substituting (\ref{eq:7}) into (\ref{eq:3}), we obtain
the radar equation relating the received signal spectrum to the medium irregularity
spectrum: 

\begin{equation}
\label{eq:9}
u(\omega )=(2\pi )^{3/2}e^{i\pi /4}\int \epsilon (\nu ,\overrightarrow{k})\frac{H(\omega -\nu ,\overrightarrow{r_{0}})g(\overrightarrow{r_{0}})V(\overrightarrow{r_{0}})(2\pi )^{3/2}}{R_{s}(\overrightarrow{r_{0}})R_{i}(\overrightarrow{r_{0}})}exp(i\varphi (\overrightarrow{r_{0}},k_{0},\overrightarrow{k}))d\overrightarrow{k}d\nu 
\end{equation}

If, instead of the coordinate system tied to wave vectors \( \overrightarrow{k} \),
 we use a system of spatial coordinates \( \overrightarrow{r_{0}} \), 
 this would amount to calculating the transition Jacobian which turns out to be 

\begin{equation}
\label{eq:10}
J=-V^{2}(\overrightarrow{r_{0}})
\end{equation}

Thus, in a spatial coordinate system, the radar equation (\ref{eq:9}) may be
written as 

\begin{equation}
\label{eq:9'}
u(\omega )=-(2\pi )^{3/2}e^{i\pi /4}\int \epsilon \left( \nu ,\overrightarrow{k}(\overrightarrow{r})\right) \frac{H(\omega -\nu ,\overrightarrow{r})g(\overrightarrow{r})}{R_{s}(\overrightarrow{r})R_{i}(\overrightarrow{r})V(\overrightarrow{r})}exp(i\varphi (\overrightarrow{r},k_{0},\overrightarrow{k}(\overrightarrow{r})))d\overrightarrow{r}d\nu 
\end{equation}

The two expressions for the received signal, (\ref{eq:9}) and (\ref{eq:9'}),
are different representations for the scattered signal written in terms of
integrals over space and over the wave vectors space, and they relate the
received signal spectrum to spectral characteristics of irregularities. 

In radar equations obtained (\ref{eq:9}),(\ref{eq:9'}) unlike the
initial equations (\ref{eq:1}),(\ref{eq:3}) the Bragg character
of scattering is emphasized (\ref{eq:6}), that is why they looks more useful
for analysis than initial ones (\ref{eq:1}),(\ref{eq:3}).

\subsection{The validity range of the expressions obtained.}

The validity range of the resulting expressions for the
scattered signal (\ref{eq:9}),(\ref{eq:9'}) is determined by the region where
we can use the first approximation of initial integral by expanding it in asymptotic
seria by the SPM technique and not taking into account next members of seria.
In accordance with SPM theory one can take into account next members,
but equations obtained is too difficult. That is why we use pretty simple criterium
\cite{Fedoruk} for estimation of first order equations validity. According
to it, next members of seria can be neglected, if the next condition is satisfied:

\begin{equation}
\label{eq:12bis}
\alpha =\begin{array}{c}
3\\
\sum \\
l=1
\end{array}\left. \frac{1}{\Delta R^{2}_{l}}\frac{\partial ^{2}}{\partial x^{2}_{l}}\frac{H(\omega -\nu ,\overrightarrow{r})g(\overrightarrow{r})}{R_{s}(\overrightarrow{r})R_{i}(\overrightarrow{r})}\right| _{\overrightarrow{r}=\overrightarrow{r}_{0}}<<1
\end{equation}

here \( l \) - cartesian coordinate \( x_{l} \) number,
\( \Delta R^{2}_{l} \) - the square of linear size in \( l \)-th coordinate
direction of the 'cophasal' region (or effective scattering volume, as we call
it (\ref{eq:8})), which makes the main contribution into the signal scattered
from exact spatial harmonic of \( \epsilon  \). 

Let us consider the limitations in the case of sounding with
the infinite signal when \( H(\omega -\nu ,\overrightarrow{r}) \) do
not depend on \( \overrightarrow{r} \). We will consider the cartesian coordinates
system linked with effective scattering volume, two basis orthes of which
lies on the plane determined by the vectors \( R_{s}(\overrightarrow{r_{0}}),R_{i}(\overrightarrow{r_{0}}) \),
one orth is parallel to the the direction transmitter-receiver. In this 
coordinate system the vector \( \Delta R^{2}_{l} \) will have the next scales:

\[
\Delta R_{l}^{2}\sim \lambda R_{i,s}\left( \frac{cos^{2}(\varphi /2)}{sin^{2}\varphi },\frac{1}{cos^{2}(\varphi /2)},1\right) ,\]

and the second derivative in (\ref{eq:12bis}) determined
by the derivatives of spatial factor \( g(\overrightarrow{r}) \) and geometrical
factor \( 1/(R_{s}(\overrightarrow{r})R_{i}(\overrightarrow{r})) \) has components
of the following order:

\[
\frac{\partial ^{2}}{\partial x^{2}_{l}}\frac{g}{R_{s}R_{i}}\sim \frac{1}{R_{i,s}^{2}}\left( \left( cos^{2}(\varphi /2)+\frac{sin\varphi }{\Delta \Theta }\right) ^{2},\left( 1+\frac{cos(\varphi /2)}{\Delta \Theta }\right) ^{2},\left( 1+\frac{1}{\Delta \Theta }\right) ^{2}\right) ,\]

where \( \Delta \Theta =\lambda /D \) - antenna pattern angle
width, and \( \varphi =\pi -\beta  \), where \( \beta =\pi -\arccos (\widehat{R_{i}}\cdot \widehat{R_{s}}) \)
- scattering angle.

Taking into account that for distances \( r \) from center
of transmitted and received antennas system to the scattered volume small 
in comparison with \( a \) (the distance between the receiver and the 
transmitter \( a=|\overrightarrow{R_{s}}-\overrightarrow{R_{i}}| \))
\( sin\varphi \approx r/a,cos(\varphi /2)\approx r/(2a),R_{i,s}\approx a \),
and for large distances \( r \) - \( sin\varphi \approx a/r,cos(\varphi /2)\approx 1,R_{i,s}\approx r \),
from the condition (\ref{eq:12bis}) we have the following validity range for
equations obtained (\ref{eq:9}),(\ref{eq:9'}):

\begin{equation}
\label{eq:13bis}
R_{i,s}>>\frac{D^{2}}{\lambda }\, \, \, \, \, \, \frac{a^{2}}{\lambda }>>r>>\sqrt{a\lambda }
\end{equation}

The first condition is coincident with validity limitations
of initial equations (\ref{eq:1}),(\ref{eq:3}) - the scattering volume must
be in the far-field range of receved and transmitted antennas, and that is why
the first condition do not make any additional limitations in comparison with
the initial ones (\ref{eq:1}),(\ref{eq:3}). The second condition requires
to exclude from analysis both forward scattering (the condition is not 
sutisfied for forward scattering when \( r\rightarrow 0 \))
and backscattering case (for monostatic experiment when \(a \approx D \) 
the condition \( a^{2}/\lambda >>r \) is not sutisfied 
and effective scattering volume (\ref{eq:8}) degenerates into infinity and 
the formulas needs appropriative modification \cite{BenrgardtPotekhin}).
Usualy in the bistatic case those conditions are sutisfied. 

For the length of the sounding signal and the receiving window
which determine the form of the kernel \( H(\omega ,\overrightarrow{r}) \),
the requirement (\ref{eq:12bis}) implies that the signal and the receiving
window change little within distances of the Fresnel radius order: 

{\par\centering \( dH(\omega ,\overrightarrow{r})/d\overrightarrow{r}\approx H(\omega ,\overrightarrow{r})\omega /c<<H(\omega ,\overrightarrow{r})/(\lambda r)^{1/2} \)\par}

This corresponds to using narrow-band signals and windows: 

\begin{equation}
\label{eq:14}
(\Delta \Omega _{o}+\Delta \Omega _{a})/c<<1/(\lambda r)^{1/2}
\end{equation}

Thus the conditions (\ref{eq:13bis}),(\ref{eq:14}) determines
the validity range of equations obtained (\ref{eq:9}),(\ref{eq:9'}).

\subsection{Selective properties of the radar equations obtained.}

Let us consider in greater detail the properties of the resulting radar equations
(\ref{eq:9}),(\ref{eq:9'}). The expressions obtained above establish a linear
relation between the scattered signal spectrum and the spatial spectrum
of dielectric density fluctuations. They clearly show selective properties of
the scattering determining the region of Fourier-harmonics which make the main
contribution to the scattering. 

The expression (\ref{eq:7}) is useful for determining the signal
scattered on some spatial harmonic. In real situations, however, the medium
involves different harmonics. Furthermore, as has been shown above, for each
spatial harmonic the greatest contribution to the scattering will be made by
the region of effective scattering (RES) whose location is determined by the
Wolf-Bragg condition (\ref{eq:6}). Also, the contribution to the scattering
from those spatial harmonics whose RES lies outside the region of beam crossing,
is small. If the beams are considered to be cones (with the angles \( \Delta \Theta _{i} \)
and \( \Delta \Theta _{s} \) for the transmitter and the receiver, respectively),
then it is possible to obtain an upper estimate (assuming that the sounding
signal and the receiving window are long enough) of the selectivity using wave
vectors from geometrical considerations. 

The spread of wave vectors involved in the scattering along directions in the
plane passing through the receiver and the transmitter is 

\begin{equation}
\label{eq:16}
\Delta \varphi =\Delta \Theta _{i}+\Delta \Theta _{s}
\end{equation}

The spread of wave vectors in absolute value is determined by the expression: 

\begin{equation}
\label{eq:17}
\Delta k=k_{mid}(\Delta \Theta _{i}+\Delta \Theta _{s}),
\end{equation}

where \( k_{mid}=2k_{0}\cos (\beta _{0}/2) \) is the wave vector corresponding
to the scattering from the the weight volume center. The wave vectors spread
along directions in the plane normal to the receiver-transmitter axis is estimated
by the formulae 

\begin{equation}
\label{eq:18}
\Delta \psi =min(\frac{R_{i}}{H}\Delta \Theta _{i},\frac{R_{s}}{H}\Delta \Theta _{s})
\end{equation}

where \( H \) is the height at which the scattering volume is located above
the ground. 

The wave vectors that make the main contribution to the scattering lie within
a region near the middle wave vector \( \overrightarrow{k_{mid}} \), corresponding
to the Bragg scattering from the center of the volume covered by the beam (\ref{eq:6}). 

A selectivity of the scattering process for the scattering by a small scatterer
was shown in \cite{Newton}; in a statistical setting of the problem,
it was estimated in \cite{Tatarsky},\cite{Isimaru}, but in a linear setting
of the scattering by an arbitrary extended scatterer this scattering process 
selectivity has not yet been established to date. Thus the resulting
radar equations (\ref{eq:9}),(\ref{eq:9'}) make it possible to establish selective
properties of the scattering, and to determine the region of wave vectors involved
in the scattering, both in direction and in absolute value. They can be used
in the analysis of scattered signals, rather than their statistical characteristics
only.

\section{The statistical radar equations.}

In this section, the radar equation is obtained for arbitrary spatial
correlation radii of irregularities, and limitations on its applicability and its
limiting cases are considered. 

Using the starting equation (\ref{eq:1}), let us develop the expression for
the scattered signal mean statistical spectral power (up to constant
factors and with the transition to a spectral representation from difference
arguments): 

\begin{equation}
\label{eq:19}
\overline{|u(\omega )|^{2}}=\int g(\overrightarrow{r})\Phi (\nu ,\overrightarrow{r},\overrightarrow{k})exp(i\varphi (\overrightarrow{k},k_{0}+K,\nu ,\overrightarrow{r}))\frac{I(\overrightarrow{k},k_{0},K,\nu )}{R_{s}R_{i}}d\overrightarrow{r}d\overrightarrow{k}\frac{dKd\nu }{(2\pi )^{4}}
\end{equation}
 where 

\begin{equation}
\label{eq:20}
I(k_{0},\overrightarrow{k},K,\nu )=\int exp(-i\varphi (\overrightarrow{k},k_{0}+K,\nu ,\overrightarrow{R}))g^{*}(\overrightarrow{R})\frac{W(\omega -\nu ,R_{s}(\overrightarrow{R})+R_{i}(\overrightarrow{R}),K)}{R_{s}(\overrightarrow{R})R_{i}(\overrightarrow{R})}d\overrightarrow{R}
\end{equation}

\begin{equation}
\label{eq:21}
\varphi (\overrightarrow{k},k_{1},\nu ,\overrightarrow{r})=\overrightarrow{k}\overrightarrow{r}+k_{1}R_{i}(\overrightarrow{r})+(k_{1}+\nu /c)R_{s}(\overrightarrow{r})
\end{equation}

\( W(\tau ,S,\Delta S)=\int o(t)a(t-S/c)o^{*}(t-\tau )a^{*}(t-\tau -(S-\Delta S)/c)dt \)
is a weighting function dependent on the signal waveform and on the receiving
window only; and \( \Phi (\tau ,\overrightarrow{r},\overrightarrow{\rho })=\overline{\epsilon (t,\overrightarrow{r})\epsilon ^{*}(t-\tau ,\overrightarrow{r}-\overrightarrow{\rho })} \)
is a stationary correlation function of dielectric permitivity fluctuations.
Its arguments are the mean statistical distance \( \overrightarrow{r} \), the
correlation radius \( \overrightarrow{\rho } \) , and the correlation time
\( \tau  \) \cite{RitovKravcovTatarsky}. It is apparent that the integrals
appearing in (\ref{eq:19}),(\ref{eq:20}) are analogous to the integral in
(\ref{eq:4}) considered above and contain a rapidly oscillating function under
the integral. We now apply the procedure described in the preceding section
to the expressions (\ref{eq:19}),(\ref{eq:20}). It is seen that the integral
in over \( \overrightarrow{R} \) (\ref{eq:20}) can be evaluated by a three-dimensional 
SPM, as done in the preceding section. The integral over \( \overrightarrow{r} \)
can be evaluated using this method on the assumption
that the dielectric permitivity fluctuations spectral density \( \Phi (\nu ,\overrightarrow{r},\overrightarrow{k}) \)
(the spatio-temporal spectrum of their correlation function) changes slowly
with \( \overrightarrow{r} \). Criteria for weak variability will be presented
below. Thus, by integrating (\ref{eq:19}),(\ref{eq:20}) over \( \overrightarrow{R} \)
and \( \overrightarrow{r} \), respectively, by the stationary-phase method,
we get: 

\begin{equation}
\label{eq:19'}
\overline{|u(\omega )|^{2}}=\int g(\overrightarrow{r_{0}})\Phi (\nu ,\overrightarrow{r_{0}},\overrightarrow{k})e^{i\varphi (\overrightarrow{k},k_{0}+K,\nu ,\overrightarrow{r_{0}})}\frac{I(\overrightarrow{k},k_{0},K,\nu )}{R_{s}(\overrightarrow{r_{0}})R_{i}(\overrightarrow{r_{0}})}V(\overrightarrow{r_{0}})(2\pi )^{3/2}e^{i\pi /4}d\overrightarrow{k}\frac{dKd\nu }{(2\pi )^{4}}
\end{equation}

\begin{equation}
\label{eq:20'}
\begin{array}{c}
I(k_{0},\overrightarrow{k},K,\nu )=-exp(-i\varphi (\overrightarrow{k},k_{0}+K,\nu ,\overrightarrow{R_{0}}))\frac{W(\omega -\nu ,R_{s}(\overrightarrow{R_{0}})+R_{i}(\overrightarrow{R_{0}}),K)}{R_{s}(\overrightarrow{R_{0}})R_{i}(\overrightarrow{R_{0}})}\\
g^{*}(\overrightarrow{R_{0}})V(\overrightarrow{R_{0}})(2\pi )^{3/2}e^{-i\pi /4}
\end{array}
\end{equation}

Furthermore, the stationary points \( \overrightarrow{r_{0}} \) and \( \overrightarrow{R_{0}} \)
depend on the wave vector \( \overrightarrow{k} \), the wave number \( K \),
and on the frequency \( \nu  \), and are defined by equations similar to the
modified Wolf-Bragg condition (\ref{eq:6}): 

\begin{equation}
\label{eq:6'}
\overrightarrow{k}=-((k_{0}+K+\nu /c)\widehat{R_{s}}(\overrightarrow{r_{0}})-(k_{0}+K)\widehat{R_{i}}(\overrightarrow{r_{0}})),
\end{equation}

whence it follows that the stationary points in (\ref{eq:19'}) and (\ref{eq:20'})
are coincident: 

\begin{equation}
\label{eq:22}
\overrightarrow{r_{0}}=\overrightarrow{R_{0}}
\end{equation}

In view of (\ref{eq:19'}),(\ref{eq:20'}),(\ref{eq:6}) and (\ref{eq:22}),
we obtain a radar equation for root-mean-square quantities in the form: 

\begin{equation}
\label{eq:23}
\overline{|u(\omega )|^{2}}=-\int \Phi (\nu ,\overrightarrow{r_{0}},\overrightarrow{k})\frac{W(\omega -\nu ,R_{s}(\overrightarrow{r_{0}})+R_{i}(\overrightarrow{r_{0}}),K)|g(\overrightarrow{r_{0}})|^{2}}{R^{2}_{s}(\overrightarrow{r_{0}})R^{2}_{i}(\overrightarrow{r_{0}})}V^{2}(\overrightarrow{r_{0}})d\overrightarrow{k}dK\frac{d\nu }{2\pi }
\end{equation}

where \( \overrightarrow{r_{0}} \) is defined by (\ref{eq:6'}). 

The radar equation obtained here relates the spectral power of the scattered
signal to the fluctuations spectral density  in the form of an integral over
the space of wave vectors. In a manner like obtaining (\ref{eq:9'}) from (\ref{eq:9}),
one can obtain the radar equation relating these two functions in terms of an
integral over space, for which purpose it suffices merely to take into account
the transition Jacobian from the coordinate system \( \overrightarrow{k} \)
to the coordinate system \( \overrightarrow{r_{0}} \) (\ref{eq:10}): 

\begin{equation}
\label{eq:24}
\overline{|u(\omega )|^{2}}=\int \Phi (\nu ,\overrightarrow{r_{0}},\overrightarrow{k}(\overrightarrow{r_{0}}))\frac{W(\omega -\nu ,R_{s}(\overrightarrow{r_{0}})+R_{i}(\overrightarrow{r_{0}}),K)|g(\overrightarrow{r_{0}})|^{2}}{R^{2}_{s}(\overrightarrow{r_{0}})R^{2}_{i}(\overrightarrow{r_{0}})}d\overrightarrow{r_{0}}dK\frac{d\nu }{2\pi }
\end{equation}

Here \( \overrightarrow{k}(\overrightarrow{r_{0}}) \) is determined by the
modified Wolf-Bragg condition (\ref{eq:6'}). The validity range of the resulting
expressions (\ref{eq:23}),(\ref{eq:24}) is constrained, in addition to the
limitations pointed out in the preceding section, by media, for which the 
fluctuations spectral density changes little with a change of the direction 
in the parameter \( \overrightarrow{r} \) by the angle \( (\lambda /r)^{1/2} \).
This implies that mean statistical properties of the medium change little in
\( \overrightarrow{r} \) within distances of the Fresnel radius order. 

Selective properties of the scattering are also pronounced in the resulting
radar equation. The beam factor \( g(\overrightarrow{r}) \) determines selection
both from wave vectors which make the main contribution to the scattering, and
from spatial regions, with the wave vector and the spatial location of the diagnosed
region being related by the condition (\ref{eq:6'}). This corresponds to the
local fulfillment of the Wolf-Bragg conditions at each point of the diagnosed
medium. The effective weight volume \( W(\omega ,S,k) \) that is determined
solely by the sounding signal and receiving window forms, determines
also the selective properties of the scattering for wave numbers, distances,
and frequencies. Its width in the spatial variable \( S \) determines the region
of transmitter-irregularity-receiver optical paths for irregularities contributing
to the scattering, while the width in the wave variable extends further the
region of fluctuations spectral density wave numbers  participating in
the scattering. The width of the wave numbers region in absolute values and
distances was considered earlier, having the order of (\ref{eq:16})-(\ref{eq:18}),
and their values are concentrated near the wave vector corresponding to the
fulfillment of the Wolf-Bragg condition for the center of the diagnosed volume.
In the frequency variable, the effective volume is convoluted with the spectral
density. For that reason, the scattered signal spectrum is mostly broader
than the frequency spectrum of the spectral density of fluctuations, and this
broadening is determined by the properties of the effective weight volume.

\subsection{Limiting cases of the radar equations obtained.}

Let us illustrate the implications of the resulting radar equation for two limiting
cases of scattering media. In the case of a time-independent isotropic medium
with a small spatial correlation radius, the irregularities spatial spectrum 
is sufficiently broad in all directions \( \overrightarrow{k} \), is
uniform in \( \overrightarrow{r} \), and has the form of a \( \delta  \)-function
in frequency \( \nu  \). For such a model of the medium, an integration can
be performed in (\ref{eq:24}) over these parameters to give a standard radar
equation for the power of the scattered signal for small spatial correlation radii 
\cite{Tatarsky},\cite{Isimaru}: 

\begin{equation}
\label{eq:25}
\overline{P}=\int \Phi (\overrightarrow{r_{0}},\overrightarrow{k}(\overrightarrow{r_{0}}))\frac{W_{1}(R_{s}(\overrightarrow{r_{0}})+R_{i}(\overrightarrow{r_{0}}))|g(\overrightarrow{r_{0}})|^{2}}{R^{2}_{s}(\overrightarrow{r_{0}})R^{2}_{i}(\overrightarrow{r_{0}})}d\overrightarrow{r_{0}},
\end{equation}

where \( W_{1}(S)=\int W(\nu ,S,K)dK\frac{d\nu }{2\pi } \). 

As a further example, we consider another limiting case which corresponds to
the case of large spatial correlation radii. Let the sounding be performed
with an infinitely long impulse with reception by an infinitely long window,
which corresponds to \( W(\nu ,S,K)=\delta (\nu )\delta (K) \). Let there exists
dielectric permittivity irregularities in the medium which have the form of 
nonstationary spatial statistically isostropic harmionic array with broad frequency 
spectrum. Then \( \Phi (\omega ,\overrightarrow{r},\overrightarrow{k})=\Phi _{0}\delta (\overrightarrow{k}-\overrightarrow{k_{1}}) \),
and the radar equation (\ref{eq:23}) becomes: 

\begin{equation}
\label{eq:26}
\overline{|u(\omega )|^{2}}=\Phi _{0}\frac{|g(\overrightarrow{r_{0}})|^{2}}{R^{2}_{s}(\overrightarrow{r_{0}})R^{2}_{i}(\overrightarrow{r_{0}})}V^{2}(\overrightarrow{r_{0}})
\end{equation}

It is evident that the spectral power of the scattered signal at each frequency
will be determined by a small region whose location is determined by the fulfillment
of an analogue for the Wolf-Bragg condition: 

{\par\centering \( \overrightarrow{k_{1}}=-((k_{0}+\omega /c)\widehat{R_{s}}(\overrightarrow{r_{0}})-k_{0}\widehat{R_{i}}(\overrightarrow{r_{0}})) \)\par}

and by dimensions the Fresnel radius (the size of the 'cophasal' region 
\( V(\overrightarrow{r}) \)) order. Thus each spectral component of the received signal will
arrive from its own point of space and with the amplitude determined by the
other terms involved in (\ref{eq:23}). Thus, by virtue of 
the radar equation proporties, the spectrum of the received signal will differ from the
spectral density frequency spectrum, and will be determined solely by the
beam. This distortion of the received signal spectrum when compared with the
medium frequency spectrum in the case of the scattering from a separate
harmonic array is not described by a standard radar equation \cite{Tatarsky},\cite{Isimaru},
and one have to consider consequent models for exact experiment
conditions for explanation of the phenomena observed experimentaly (for example,
for radioacoustic sounding this leads to the signifficant difference between
expected doppler shift and observed frequency shift of received signal \cite{KonTastarski}).

\section{Conclusion}

In this paper we have outlined the method for obtaining the relation between 
scattered signal and spectral characteristics of the medium, which implies 
essentially the transition to a consideration of the scattering on medium 
spatial  harmonics and to the the stationary-phase method implementation 
for calculating the contributions from separate harmonics. 

By using the proposed method we have obtained two equivalent radar equations relating
the Fourier-spectrum of the scattered signal to the spatial Fourier-spectrum
of dielectric permitivity fluctuations (\ref{eq:9}),(\ref{eq:9'}) for the
case where the receiver and the transmitter are not in the far-field range of
the scattering volume. These equations show explicitly the selective character
of the scattering process: the main contribution to the scattering is made by
spatial harmonics of the medium, for which conditions similar to the Wolf-Bragg
conditions are satisfied (\ref{eq:6}). The validity range of the expression
obtained in this study is virtually coincident with that of the starting expression
in a spatio-temporal representation (\ref{eq:1}) provided that sufficiently
narrow-band sounding signals are used (\ref{eq:14}) and the cases of exactly
backward (considered in \cite{BenrgardtPotekhin}) and
exactly forward soundings are excluded (\ref{eq:13bis}). Thus the resulting
expressions (\ref{eq:9}),(\ref{eq:9'}) are an analogue for (\ref{eq:1}) in
a spectral region in most bistatic experiments on remote probing
of media. The radar equation obtained in this study is a generalization of formulas
for small scatterers \cite{Newton} to scatteres of an arbitrary
size. In problems of media diagnostics, equations (\ref{eq:9}),(\ref{eq:9'})
can be used to analyze scattering signals as such, rather than their statistical
characteristics alone. 

The proposed method has been used to obtain two equivalent, statistical radar
equations for arbitrary radii of spatial correlation (\ref{eq:23}),(\ref{eq:24})
which hold true for media whose mean statistical parameters change little within
distances of the Fresnel radius order. This limitation usually
true for measurements of real media, because mean statistical parameters of
the medium (scattering section, drift velocity, etc.) usually vary smoothly
throughout the diagnosed volume. 

It has been shown that these radar equations (\ref{eq:23}),(\ref{eq:24}) can
be useful for obtaining both a standard radar equation for small spatial 
correlation radii of irregularities (\ref{eq:25}) and radar equations for other models
of scattering media. 

The statistical radar equations obtained (\ref{eq:23}),(\ref{eq:24})
one should use instead of standart in the cases where spatial correlation radii
are compared to or greater than Fresnel radius in some direction. The situations
can arise, for example, in case of scattering from ionospheric irregularities
elongated with Earths magnetic field and in case of scattering from anisotropic
air turbulence and do arise in case of atmosphere radioacoustic sounding.
The non-statistical radar equations obtained (\ref{eq:9}),(\ref{eq:9'})
may be useful for analysis of scattered signal without averaging.

\newpage
\end{document}